\documentclass[a4paper, twoside, 11pt]{article}
\usepackage{natbib}

\usepackage[a4paper, margin=1in]{geometry}
\usepackage[latin1]{inputenc}
\usepackage[T1]{fontenc}
\usepackage[english]{babel}
\usepackage{textcomp} 
\usepackage{graphicx}
\DeclareGraphicsExtensions{.jpg,.mps,.pdf,.png,.gif}
\usepackage[french]{varioref} 
\usepackage[]{amsmath,amssymb,amsfonts}

\usepackage{float}
\usepackage{bbm}
\usepackage{bbold}
\usepackage{amsthm}

\newcounter{magicrownumbers}

\usepackage{ifsym}
\usepackage{mathptmx}
\usepackage{natbib}
\usepackage{helvet}
\usepackage{courier}
\usepackage{dsfont}
\usepackage{color,calc}  
  \usepackage{pgf}
  \usepackage{graphicx}
  \usepackage{multicol}
  \usepackage{amsmath,amssymb}
  \usepackage{bbm} 
  \usepackage{multicol} 
\usepackage{color, colortbl}
\usepackage{lscape}
  
\usepackage{type1cm}         

\usepackage{makeidx}         
\usepackage{graphicx}        
\usepackage{multicol}        
\usepackage[bottom]{footmisc}

\definecolor{blue1}{rgb}{0,0,0.5451} 
  \definecolor{blue2}{rgb}{0.2,0.4,0.65}
\definecolor{DarkBlue}{rgb}{0,0,0.5451}  
\definecolor{darkred}{rgb}{0.75,0.0,0.0}
\definecolor{darkgreen}{rgb}{0.0,0.6,0.0}
\definecolor{darkblue}{rgb}{0.0,0.0,0.6}
\definecolor{darkcyan}{rgb}{0.0,0.6,0.6}
\definecolor{darkmagenta}{rgb}{0.6,0.0,0.6}
\definecolor{darkyellow}{rgb}{0.6,0.6,0.0}
\definecolor{lightred}{rgb}{1.0,0.9,0.9}
\definecolor{lightgreen}{rgb}{0.9,1.0,0.9}
\definecolor{lightblue}{rgb}{0.9,0.9,1.0}
\definecolor{lightcyan}{rgb}{0.8,1.0,1.0}
\definecolor{lightmagenta}{rgb}{1.0,0.8,1.0}
\definecolor{lightyellow}{rgb}{1.0,1.0,0.8}
\definecolor{paleyellow}{rgb}{1.00,1.0,0.80}
\definecolor{amber}{rgb}{1.0,0.8,0.0}
\definecolor{darkamber}{rgb}{1.0,0.5,0.0} 
\definecolor{Gray}{gray}{0.9}

\definecolor{gray}{cmyk}{0,0.1,0.2,0.3,0.4,0.5,0.6.0.7,0.8,0.9}

\newcommand{\Ben}{\begin{enumerate}}
\newcommand{\Een}{\end{enumerate}}
\newcommand{\Bit}{\begin{itemize}}
\newcommand{\Eit}{\end{itemize}}
\newcommand{\Beq}{\begin{equation}}
\newcommand{\Eeq}{\end{equation}}
\newcommand{\Ba}{\begin{align*}}
\newcommand{\Ea}{\end{align*}}
\newcommand{\Mb}{\mathbf}

\title{
A frailty-contagion model for  multi-site hourly precipitation driven by atmospheric covariates
}

\begin{document}
\author{
Erwan Koch\footnote{ISFA, CREST and ETH Zurich (Department of Mathematics, RiskLab). erwan.koch@math.ethz.ch 
}
\and Philippe Naveau \footnote{LSCE (CNRS). philippe.naveau@lsce.ipsl.fr}
}

\date{}

\maketitle


%
%
\begin{abstract}
 Accurate stochastic simulations of hourly precipitation are needed for impact studies at local spatial scales. 
Statistically, hourly precipitation data represent a difficult challenge. They are non-negative, skewed, heavy tailed, contain a lot of zeros (dry hours) and they have complex temporal structures (e.g., long persistence of dry episodes).   
Inspired by  frailty-contagion approaches used in finance and insurance, we propose a multi-site precipitation simulator  that, given  appropriate regional atmospheric variables, can  simultaneously handle dry events and heavy rainfall periods. 
One advantage of our model is its conceptual simplicity  in its dynamical structure. 
In particular, the temporal variability  is represented by a  common   factor  based on a few  classical atmospheric  covariates  like temperatures, pressures and others. 
Our inference approach is tested on simulated data and applied on measurements made in the northern part of French Brittany. 

\medskip
 
\end{abstract}

\newpage

\section{Introduction}

Stochastic weather generators (WG)  are statistical models that aim  at simulating  quickly and  realistically  random  sequences of atmospheric variables like temperatures, precipitation and wind speeds. 
Historically,   weather generators started in  hydrological sciences  in the sixties and seventies. 
In  \citeyear{Gabriel62}, \citeauthor{Gabriel62} proposed a Markov model for daily precipitation occurrences.  
Since then, a   strong research effort   in modeling precipitation distributions has sustained in the hydrological and statistical communities. 
This attention towards precipitation WGs   can be explained by, at least,   two  different reasons. 
These stochastic simulations can be used in  assessment studies,  especially these  linked to   water resources   managements. 
As one of the initial drivers in impact studies, simulated times series of precipitation can play a fundamental role in   exploring  some part  of  the {sensitivity} 
of floods, erosion and  crops models.  
 The second reason, very attractive to applied statisticians, is that modeling accurately precipitation distributions in space and time has always been  an  intriguing   mathematical challenge.  
From a probabilistic point of view, a sequence of   precipitation  mixes two types of events (dry or wet) and the rainfall intensity represents a strong departure from 
the Gaussian territory (positive, skewed and sometime heavy tailed).  Spatially and temporally, events can be strongly correlated within dry or wet episodes. 
Still today,  the spatio-temporal dynamics of precipitation is difficult to model statistically. 
Numerous elaborate  parametric approaches  have  been proposed in recent years \cite[see, e.g.][]{Furrer08,lennartsson08,Vrac07,Allard12} (non-parametric approaches also exist but we will not discuss them here). 

\begin{figure}[h]
\center
\includegraphics[scale=.5]{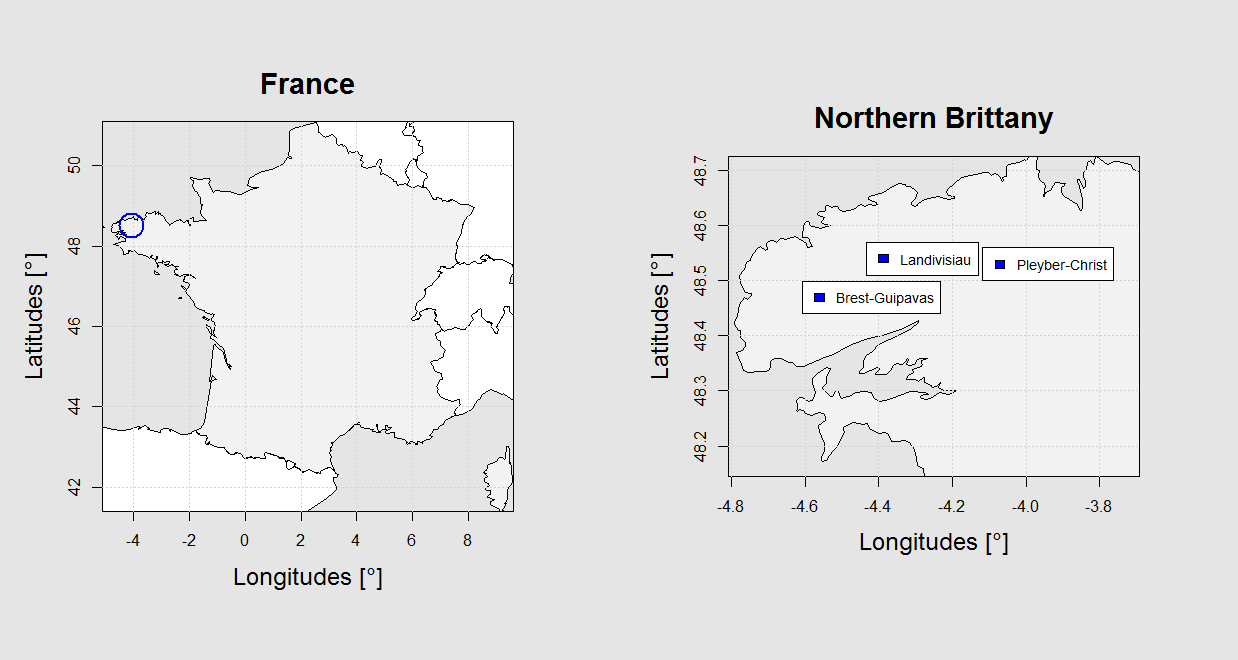}
\caption{Three weather stations in northern Brittany  (France) with hourly winter precipitation records from  {1995} to {2011}. }
\label{Stations}
\end{figure}

From a probabilistic  point of view, most of daily precipitation generators  decouple   the occurrence   and  intensity processes. 
For example, \cite{Kleiber12} first generated  spatially and temporally correlated rainfall  occurrences and then, at locations with positive precipitation, simulated their intensities. 
This    strategy  is classical in the WG literature   \cite[see, e.g.][]{Katz77,Richardson81}.
It is   mathematically convenient, in terms of inference,  to frame the estimation scheme into a  clear two step algorithm.  
But the hypothesis that the occurrence process can be simulated before the intensity process, i.e. independently of  past, present or future rainfall amounts,  could be challenged, especially at the hourly scale. 
If very  large (small) rainfall amounts have been observed at time $t$, it seems more likely that a wet (dry) hour 
will follow at time $t+1$. 
To quantify this naive reasoning, we need to  introduce the   example that will lead all our discussions in this article. 
 We will work with winter (DJF) hourly precipitation recorded in the northern part of Brittany in France at  three stations from {1995} to {2011}, see Figure \ref{Stations}.
In a nutshell, winter precipitation in this region are generally due to large scale perturbations coming from the Atlantic ocean.

Coming back to the question of simulating the occurrence process independently  of the intensity, 
Table \ref{Occurrence probability given intensity winter} shows that the  probability of rainfall occurrence at time $t+1$ increases in  function of the rainfall intensity at date $t$ for each  station. 
\begin{table}[h!]
\center
\begin{tabular}{|c|ccc|}
\hline
\multicolumn{1}{|c|}{\bf  } & \multicolumn{3}{c|}{ Weather stations in Fig. \ref{Stations} }\\
\hline
{Intensity range (mm)} & {Brest-Guipavas} & {Landivisiau} & {Pleyber-Christ} \\ 
\hline
\rowcolor{Gray}  0.2-0.8 & 0.63 & 0.65 & 0.68 \\ 
0.8-1.4 & 0.74 & 0.75 & 0.79 \\ 
\rowcolor{Gray} 1.4-2 	& 0.81 & 0.82 & 0.82 \\ 
2-4     & 0.90 & 0.91 & 0.91 \\ 
\rowcolor{Gray}  > 4 	& 0.94 & 0.94 & 0.94 \\ 
\hline

{Chi-Square test p-value } & $8.40 \times 10^{-64}$ & $4.57 \times 10^{-59}$ & $4.39 \times 10^{-43}$ \\ 
\hline
\end{tabular}
\caption{Rain occurrence probability  at time $t+1$ given the rainfall  intensity at $t$. 
Each row  represents a specific range of hourly rainfall  intensity.  
The Chi-Square   p-values test the independence hypothesis.} 
\label{Occurrence probability given intensity winter}
\end{table}
 To couple the occurrence and intensity processes, a possible approach, see e.g. \cite{Ailliot09} and \cite{Kleiber12} who discussed this issue on p.4,
 is to first  generate a Gaussian random field, say $Z_t$. Secondly, a censoring mechanism applied to this field produces rainfall  occurrences,  i.e. wet events occur  whenever  $Z_t$ is above a given threshold. Then,   rainfall intensities are obtained  by   transforming   the Gaussian values   above the threshold into   positive  quantities \cite[e.g., see][]{Allard13}. 
 The resulting rainfall intensity  distribution depends on the choice of the transform function  that may have different    flavors \cite[see, e.g.][]{Wilks98,Furrer08,lennartsson08,Vrac07,Allard12}. 
 Concerning the correlation structure, the hidden process $Z_t$, by construction,   drives both the intensity and occurrence processes that  {have} 
 become dependent. 
Another   strategy \citep{Serinaldi14} is to directly fit a   discrete-continuous at-site mixture distribution with two components (zero  or rainfall amount). 
 The parameters of these two components are then modelled  by spatio-temporal processes (General Additive Model (GAM) or General Linear Model (GLM)).

 This   idea of working with an unobserved ``seed" process is mathematically appealing but it adds some difficulties.  In terms of statistical inference, it makes the estimation difficult. Concerning the interpretation, the latent  process and its associated parameters are not necessary simple to explain. 
Although it has been a classical strategy to introduce hidden quantities in weather generators, for example see   the abundant WG  literature concerning hidden Markov structures \cite[see reviews like ][]{WilksWilby99,srikanthan01,Ailliot14},  
it would be a clear  improvement, in terms of interpretation, if multi-site precipitation generators  could bypass the hypothesis of hidden processes and directly model  raw observations. 
This implies that additional information characterizing the dynamical structure of the weather system has to be integrated in a different way.

A well-maintained weather station recording hourly precipitation is rarely climatologically  isolated in the sense that atmospheric variables like temperatures, pressures or  others should be available in the same region. 
Most likely, the weather station itself may monitor this type of data.
In this context, our plan is to develop the following two simple ideas in order to propose  a new multi-site precipitation  generator:
\begin{enumerate}
\item[(A)] the main driver of current precipitation   at a single location are   precipitation recorded one hour early  at all sites,   
\item[(B)]  the second  source of information comes from   atmospheric variables like temperatures  and others  that jointly impact all sites by   dynamically driving  the   variability of the residuals obtained from step (A).
\end{enumerate}
Although  resembling to  existing concepts, this two step roadmap differs from the aforementioned studies by a few but important points. 
We will not  represent  the precipitation range by a finite  mixture of  {distributions}
  based on  numerous weather types, each one  representing    different atmospheric conditions. 
Neither an  hidden Gaussian process nor a   function to transform a Gaussian variable into a Gamma type  rainfall intensity are needed.

Our point (A) is not novel and it corresponds to the well-known idea of an auto-regressive process \cite[see, for example the Markov chain in][]{Katz77}. 
Here, we will use a basic multivariate auto-regressive representation \cite[e.g., see][]{Davis14,Grimaldi05}. 
Point (B), modeling the residuals as a dynamical function of atmospheric variables, could be considered  as more innovative in the WG literature. 
It is closely linked to the research developed in the statistical downscaling literature \cite[e.g., see ][]{maraun2010,wilks2010use, wilks2012stochastic}. 
Downscaling approaches have mainly focused {red}{on}
 how to make the connection between circulation patterns and local atmospheric variables at the daily scale \cite[e.g., see ][]{Vrac07a,Flecher10}. 
Besides a different  temporal scale (hourly in our case),  our ``spatial" component   differs  from downscaling. Our ``common" signal,  ordinary  atmospheric variables   averaged over our three sites in northern Brittany, corresponds to an information shared by all the sites, but 
it does not  represent a regional feature   over hundred of kilometers.  
The common point with downscaling  is rather the idea that precipitation simulators  performances can be improved if some appropriate explanatory  variables can be  injected in the statistical model. 

The most interesting point of our article resides in  combining      ideas (A) and (B). From this coupling, a very simple model can handle  precipitation in northern Brittany at the hourly scale, not  only in terms of moderate rainfalls but also in terms of intense dry periods and even heavy precipitation.  
Before describing in detail our model, see Section \ref{sec: model}, we need to say that   our modeling strategy has been inspired by tools coming   from  two   research fields not directly linked to  geosciences:  systemic risk in  insurance  and epidemic models.

Analyzing systemic risk in the financial system involves modeling the dependence between institutions. A first kind of dependence comes from   common risk factors. These are called systematic factors, or frailties. For instance, the risk of life insurance contracts depends on the general uncertain increase of human lifetime, called longevity risk. 
As highlighted by the regulation since 2008,  the second main source of dependence stems from contagion phenomenon between institutions. The contagion effect is due to the interconnections between banks and insurance companies by means of their debt and shares participations. Typically, the failure of a company
will imply losses for its lenders, and maybe the failure of some of them. Then failure of some lenders of the lenders  can occur, and so on. 
Originally,  contagion models in finance come from the epidemic model introduced by 
\cite{bailey1953total, bailey1957mathematical}, \cite{kendall1956deterministic} and reintroduced in the so-called infectious model used in a static framework by \cite{davis2001modelling}. Specifications including both frailty and contagion effects are introduced among others in \cite{frey2003interacting}, \cite{giesecke2004cyclical, giesecke2006credit}, \cite{azizpour2008self}, \cite{gagliardini2013correlated}.

Coming back to   hourly  precipitation data, our weather system in Brittany also  contains both a common factor (frailty term) and a contagion term. The analogy with the financial system is the following. 
Each weather  station   corresponds to an institution. Our common factor represents large scale conditions that are, in some sense, exogenous to our system. 
For example, a storm front coming from the Altantic ocean is not directly  {produced}
  by the weather system observed in Brittany.  It is   similar to an exogenous shock affecting the whole financial system. Then once large scale processes are set then   local physical processes are involved. 
These processes depend on the local characteristics. For instance, presence of mountains or flows can originate a privilegiate direction for thunder-storms propagation. It is some kind of contagion from one site to another. These local interactions between sites are the equivalent {of}
 cross-holdings at the origin of contagion in the financial system. 
Contrary to the usual assumptions used in finance, the common factor in climatology can be  observable,  covariates as temperatures, winds,  pressures, and so on are also recorded at the weather stations and they represent valuable information that we have to take into account. Another link with financial statistics resides in the assumption that rainfall variability will be modelled by an heteroscedastic variance, a classical feature in econometrics. See e.g. the ARCH/GARCH model literature. 

The remaining of the paper is organized as follows. Our heteroscedastic multi-site rainfall generator is detailed in Section \ref{Chaprainfall_Sec_Model}. Then Section \ref{Sec_Estimation_Simulation} presents our inference method  based on maximum likelihood. Section \ref{Sec_Applications} assesses our model's performance on simulations and summarizes our case study results  dealing with   the northern part of Brittany. Finally a discussion is given in Section \ref{Chaprainfall_Sec_Conclusion}.

\section{A heteroscedastic  multi-site rainfall generator}\label{sec: model}
\label{Chaprainfall_Sec_Model}

Denote $M$ the number of weather stations and   $P_{m,t}$ the   precipitation amount at station $m$  recorded during    the $t^{th}$ hour. 
Our multi-site model based on (A) and (B)  is defined as follows 
\Beq
\label{General_Model}
P_{m,t}=
\left \{
\begin{array}{cc}
{\bf B}_m^{'} {\bf P}_{t-1} + \epsilon_{m,t} & \mbox{ if } {\bf B}_m^{'} {\bf P}_{t-1} + \epsilon_{m,t} \geq u, \\
0 & \mbox{ if }  {\bf B}_m^{'} {\bf P}_{t-1} + \epsilon_{m,t} < u,
\end{array}
\right.
\Eeq
where the  scalar product 
$$
{\bf B}_m^{'} {\bf P}_{t-1} = \beta_{m,1} P_{1,t-1} + \dots + {\beta_{m,M}} 
P_{M,t-1}
$$
between the vector ${\bf P}_{t-1} =( P_{1,t-1}, \dots, P_{M,t-1} )^{'}$   and ${\bf B}_m = (\beta_{m,1}, \dots, \beta_{m,M})^{'}$ 
represents  the multivariate auto-regressive  vector of order one that captures 
 the dynamical  local-scale effect, i.e. how  the neighborhood behavior  affects  station $m$ one hour  later. 
 The $M \times M$ unknown auto-regressive coefficients $\beta_{i,j}$ can be  concatenated into  a matrix, say  $\bf B$,  composed of  the  {rows}
  ${{\bf B}_m^{'}}$.
  For our Brittany example,   large scale perturbations coming from the Atlantic ocean should  make the matrix $\bf B$  asymmetric. 
The Brest weather station should influence the two stations eastward, while the converse should not be true.

The random variable $\epsilon_{m,t}$ in (\ref{General_Model}) corresponds to  the local variability  driven by   a few  atmospheric variables  and $\epsilon_{m,t}$ is simply modeled by a  sequence of zero-mean  Gaussian independent random variables. 
The non-stationarity comes from the standard deviation of  $\epsilon_{m,t}$, $\sigma_t>0$,  that  varies with time $t$  in a log linear fashion 
 \Beq
\label{eq: var esp}
\ln \sigma_t =\Mb{\theta}^{'} \Mb{F}_t, 
\Eeq
 where $\Mb{F}_t$ represents a  vector of $d$ atmospheric explanatory variables at time $t$.  
 The vector of scalars  $\Mb{\theta}$  corresponds to the unknown regression terms of the  log-linear model in Equation (\ref{eq: var esp}).
 Concerning the spatio-temporal features  of Equation (\ref{eq: var esp}),  neither $\theta$ and $ \Mb{F}_t$   change from site to site, they only contain a  spatially pooled  information shared by all sites,  and only  $ \Mb{F}_t$ varies with time.  
 With a financial vocabulary, an economist to describe (\ref{General_Model}) and (\ref{eq: var esp})   will speak of a heteroscedastic model with a volatility  driven by the  "frailty term" (exogenous common factor) $\Mb{F}_t$.

To ensure non-negative precipitation and  produce dry events, Equation (\ref{General_Model}) contains  the condition that  rainfall only occur if the quantity ${\bf B}_m^{'} {\bf P}_{t-1} + \epsilon_{m,t} $ is  greater than some positive fixed threshold $u$. This could be loosely  interpreted as saying  that the rainfall generated  at time ${t}$
 , ${\bf B}_m^{'} {\bf P}_{t-1} + \epsilon_{m,t}$, has to {be}
  large enough to be recorded by the $m^{th}$ weather station.  
This view emphasizes the fundamental role  of having a temporally varying $\sigma_t$.

Imagine that all stations are dry  at time $t-1$. All elements of the vector  ${\bf P}_{t-1}$ would  then be equal to zero. 
In this case, the only way to produce large (small) rainfall at time $t$ is to randomly generate a  large (small)  $\epsilon_{m,t}$. 
This can be done by choosing a large (small) volatility, $\sigma_t$, or equivalently a  large (small) frailty term $     \Mb{F}_t $. 
In other words,   large scale conditions  are captured by  $     \Mb{F}_t $ 
that drives the volatility of the system, and ultimately   {drives}
  precipitation occurrences and strengths. 
This implies that heavy rainfall behavior, as well as dry period persistence, can  be  reproduced only if  an adequate vector  $     \Mb{F}_t $ is chosen. 

Compared to \cite{Kleiber12}, the dependence between occurrence and intensity is directly built in Equation (\ref{General_Model}). 
If it had rained  a lot at time $t-1$, then ${\bf P}_{t-1}$ is large, and it is very likely to have a wet hour
at time $t$ because of (\ref{General_Model}). In the opposite case, if ${\bf P}_{t-1}$ is equal to zero, it is rather unlikely to have a wet hour at time $t$, unless the volatility is large. So,  current rainfall occurrences   are correlated with the intensity value observed one hour early. 

%
%

Concerning the rain intensity distribution  itself,  our hypothesis that all $\epsilon_{m,t}$ are normally distributed does not imply 
that our  simulated  rainfall will follow  a truncated Gaussian density, and consequently this does not mean that  heavy tailed behavior cannot be simulated.
The non-stationarity of $\sigma_t$ explains this phenomenon. At any time, $\sigma_t$ can take a large value and therefore simulated precipitation  resembles to a complex infinite mixture of truncated Gaussian random variables with a wide range of standard deviation.  
A consequence of this is that we don't need, at least in our example, to apply a power transform to go back and forth between the Gaussian world and the rainfall values like in \cite{Allard13} and \cite{Ailliot09}. 
Another reason is that $\epsilon_{m,t}$ does not represent   raw precipitation but 	a type of increment between two consecutive hours, see Equation (\ref{General_Model}).

As emphasized in the previous paragraphs, the choice of  $\Mb{F}_t$ is paramount   in the overall capacity of our model to accurately simulate hourly precipitation.
In order to limit overfitting, the dimension of $d$ should not be too high and, to make our model useful,  
the type of covariates within $\Mb{F}_t$
should be easy to obtain in most rainfall applications. 
For example, the components of $\Mb{F}_t$  in our northern Brittany case are hourly temperatures,   pressures at the sea level,  and  humidity, respectively.  
For each atmospheric variable, we spatially average   hourly values recorded  over our three sites to get $\Mb{F}_t$. 
Before closing this section, we would like to emphasize that the choice of the threshold $u$ plays an important role in accurately modeling  the length of dry periods. 
The coming two sections will illustrate this point.

\section{Inference}
\label{Sec_Estimation_Simulation}
Given the vector $\Mb{F}_t$ and the threshold $u$, our inference scheme is  based on maximizing  the likelihood (ML) of model (\ref{General_Model}) with respect to   the matrix of auto-regressive parameters ${\bf B}$ and the regression coefficients $(\theta_0, \dots, \theta_d)$ where $\theta_0$ represents the intercept. 
The log-likelihood function denoted  $L_u({\bf B}, \theta_0, \dots, \theta_d)$ for a given $u$   can be written as (see the proof in the Appendix)    
\begin{align}
&L_u({\bf B}, \theta_0, \dots, \theta_d) \nonumber
\\& = \sum_{t=2}^T \sum_{m=1}^M \left\{ 
{\bf I}_{ \{ P_{m,t} \geq u  \} } \ln \Bigg[ \frac{1}{\sigma_t} \phi \left( \frac{P_{m,t} - {\bf B}_m^{'} {\bf P}_{t-1}}{\sigma_t} \right) \Bigg] + {\bf I}_{ \{P_{m,t} = 0 \} } \ln \Bigg[ \Phi \left( \frac{u - {\bf B}_m^{'} {\bf P}_{t-1}}{\sigma_t} \right)  \Bigg]
\right\} \label{eq: like},
\end{align}
where  $\Phi (.)$ and $\phi (.)$ represent  the   cumulative distribution function and the probability distribution function of the standardized normal distribution, respectively. 
The vector  $(\theta_0, \dots, \theta_d)^{'}$ appears throughout $\sigma_t$, see Equation (\ref{eq: var esp}). 
The indicator function ${\bf I}_{ \{ P_{m,t} \geq u \} }$, equal to one  if $ P_{m,t} \geq u$ and zero otherwise,  corresponds to our condition in (\ref{General_Model}) that generates either a dry or wet hour at station $m$. 
These indicator functions and the absence of a closed form  for $\Phi(.)$ {make}
 the derivation of explicit ML estimates impossible, these values can only be obtained   numerically.
Our confidence intervals are derived by using  a profile log-likelihood approach \citep{Yudi01}.

The above ML approach assumed that the threshold $u$ has been correctly  chosen. 
As our  weather station precision is $0.2$ mm, the scalar $u$ is iteratively  set  to small values close to this instrument precision limit, say    $0.2, 0.3, 0.4, 0.5, 0.6$ and $0.7$.    
To find the optimal $u$ from this set, we arbitrarily chose a first guest, say $.5$ mm, and implement our ML inference that provides estimates for $({\bf B}, \theta_0, \dots, \theta_d)$. 
For this set of ML estimates, we can simulate hourly precipitation for different values of $u$ and then, choose the 
   value of $u$ minimizing the difference between the observed mean length of dry periods and the mean length of dry periods of the simulated precipitation. 
  This process can be repeated with the new value $u$ and we stop when the estimated threshold remains unchanged. 
  To conclude on the choice of $u$, we note that, by construction, model (\ref{General_Model})  will never simulate values above  zero but smaller than $u$. 
Hence, if our fixed $u$  is little bit greater than the instrument precision of $.2$ mm, then very small but positive precipitation in Brittany  {are not considered}
in our ML optimization scheme, see Equation (\ref{eq: like}). 
This technical detail does not play an important role  in our overall modeling strategy  but allowing $u$ to be greater than $.2$ mm improves substantially   our capacity to reproduce dry period lengths.

\section{Application}
\label{Sec_Applications}

\begin{figure}[H]
\center
\includegraphics[scale=0.4]{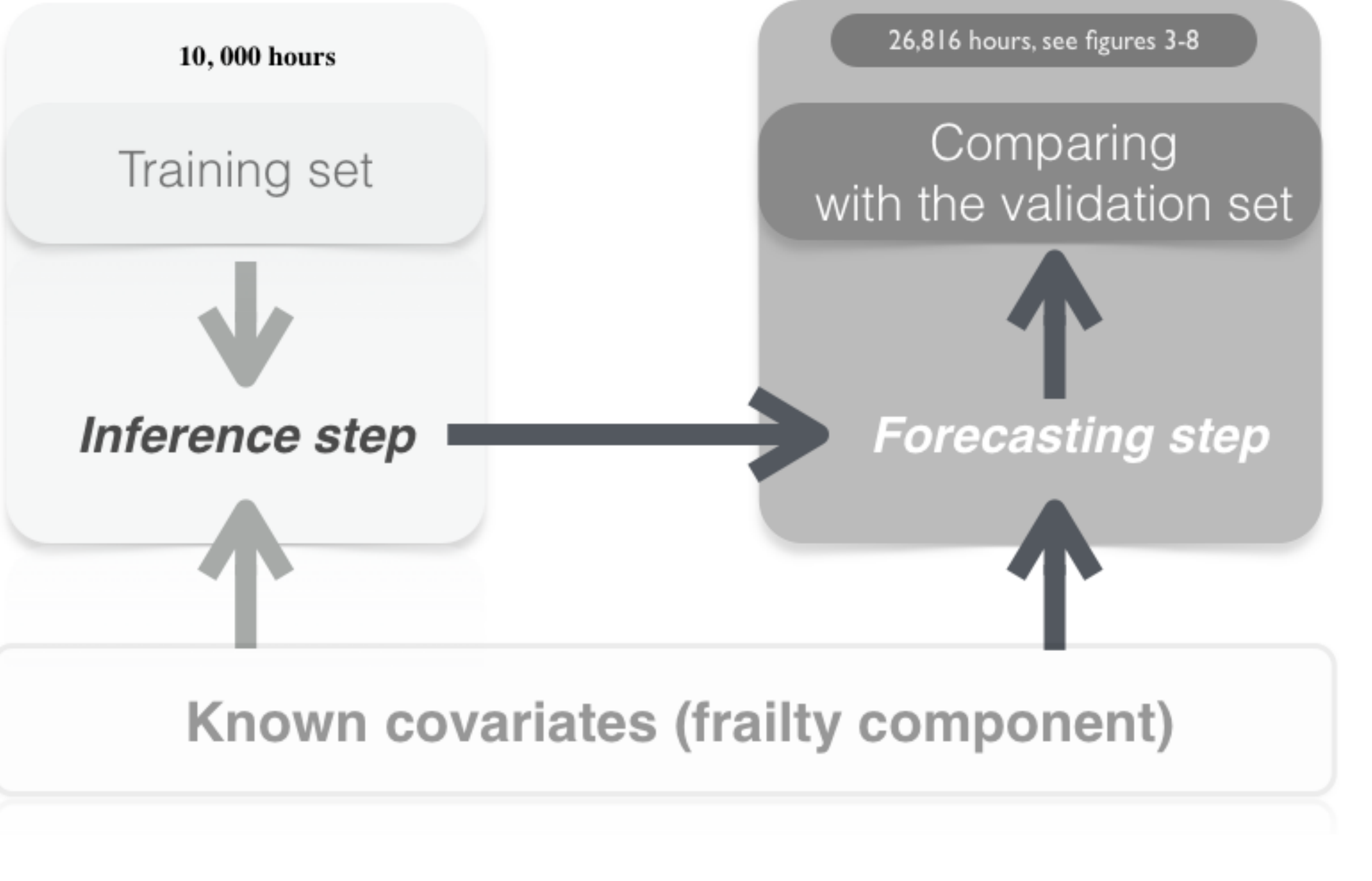}
\caption{
Inference and validation scheme: a learning set of   $10,000$ hours is used to infer the parameters of our model. 
Then, assuming that the covariates are known in the future, we predict during the next $26,816$ hours and compare forecasted rainfall (so-called out-of-sample predictions) with   precipitation  recorded  in Brittany.
A few visual summaries of this forecast are shown in figures 3-8.}
\label{fig: design}
\end{figure}

For our simulations and Brittany example,  a learning set of   $10,000$ hours and a validation set of $26,816$ hours  are used  to assess the performance of our estimation scheme. 
Figure \ref{fig: design} summarizes our inference and validation scheme.

\subsection{Simulations}
To test our inference scheme, we simulate $100$ independent replicas for our model (\ref{General_Model}) with    parameters that   are chosen to be similar to  our Brittany example, see the first column of Table \ref{Performance_Estimation}.
For two sample sizes ($100$ and $1000$), 
the parameter  estimates are derived by a classical {ML}
 approach based on Equation (\ref{eq: like}) and indicates that inferring 13 parameters with a small sample may be done with caution. 
The empirical mean, standard deviation and relative error derived from our    100 replicas are shown in Table \ref{Performance_Estimation}.
As expected, the sample length plays an important role in the inference. 
For a size of $100$, the estimation can be difficult. In contrast, working with a sample size of $1000$ provides accurate estimates. 
In practice, the sample size {for}
 hourly data  is often long,  i.e. more than $1000$ hours. For our Brittany example, we have inferred our parameters with $10,000$ observations. 
In our simulations, the inference accuracy with a sample size of $10,000$ points is  slightly better to the one with $1,000$ observations, except for $\theta_0=30.63$ which is much improved (the bias and stdev become $-0.005$ and  $0.59$, respectively).

\begin{table}[h]
\center
\begin{tabular}{|c|ccc|ccc|}
\hline
  Sample size &  \multicolumn{3}{ |c| }{100} &  \multicolumn{3}{ |c| }{1000}  \\ 
\hline\hline
 \rowcolor{Gray}  Parameter values & Bias & Stdev & Relative error & Bias & Stdev & Relative error \\ 
\hline
\hline
$\beta_{11}=0.65$   &   -2.08   &   17.71   &   -3.19   &   0.00   &   0.05   &   0.00 \\
$\beta_{12}=-0.08$   &   -8.74   &   22.97   &   114.37   &   -0.01   &   0.06   &   0.07 \\
$\beta_{13}=0.11$    &   -13.30   &   30.62   &   -122.71   &   0.00   &   0.06   &   0.03 \\
$\beta_{21}=0.47$   &   -3.44   &   17.94   &   -7.37   &   0.01   &   0.05   &   0.02 \\
$\beta_{22}=0.25$   &   -5.57   &   13.83   &   -22.36   &   0.00   &   0.04   &   -0.01 \\
$\beta_{23}=0.02$   &   -7.76   &   20.43   &   -435.58   &   0.00   &   0.06   &   -0.09 \\
$\beta_{31}=0.22$    &   -8.64   &   27.02   &   -39.95   &   0.00   &   0.05   &   -0.01 \\
$\beta_{32}=0.10$   &   -7.17   &   21.43   &   -75.40   &   0.00   &   0.05   &   0.02 \\
$\beta_{33}=0.36$    &   -8.18   &   24.75   &   -22.65   &   -0.01   &   0.05   &   -0.03 \\ \hline \hline
$\theta_0=30.63$   &   -12.36   &   146.13   &   -0.40   &   0.19   &   1.54   &   0.01 \\
$\theta_1=0.07$   &   0.07   &   0.36   &      0.96   &   0.00   &   0.01   &   -0.01 \\
$\theta_2=0.03$   &   -0.06   &   0.14   &   -0.27   &   -0.07   &   0.00   &   0.01 \\
$\theta_3=0.03$   &   0.03   &   0.08   &        0.92   &   0.00   &   0.00   &   0.02 \\
\hline
\end{tabular}
\caption{
Inference assessment by simulations for two sample sizes of $100$ and  $1000$. 
The empirical bias, standard deviation and relative error are derived from  100 independent replicas simulated from (\ref{General_Model}) with    parameters given in the first column.
 }
\label{Performance_Estimation}
\end{table}

\subsection{Hourly precipitation in northern Brittany}
The estimated  {auto-regressive coefficients}
 $\beta_{ij}$ of the matrix $\bf B$ with the 95\% confidence intervals and coverage probability 
are displayed in  Table \ref{Tab MAR}. 
The first column makes it  clear that the weather station of Brest has a strong influence in the   two  eastward  stations. 
In particular, the station of Landivisiau has a larger coefficient with Brest ($.47$) than with itself ($.25$).
 In contrast, the two stations of Landivisiau and  Pleyber-Christ, which are close to each other,  do not have an impact on Brest. 
 This pattern corresponds to the expected behavior for this region where rainfall comes from the Atlantic side, see Figure \ref{Stations}.
 This implies that Brest would be the first one to be impacted by a westerly front, and then  the two eastward stations will be hit by the storm later on. 
 \begin{table}[h]
\center
\begin{tabular}{|c|ccc|}
\hline
{\bf Matrix B} &  {Brest-Guipavas} & {Landivisiau} & {Pleyber-Christ} \\ 
\hline
\rowcolor{Gray}  {Brest-Guipavas}  & 0.65 &  -0.08 & 0.11 \\ 
\rowcolor{Gray}   				  & [0.59 ; 0.74] (87)  & [-0.15  ; 0.01] (86) & [0.02 ; 0.19]  (89) \\ 
Landivisiau & 0.47 & 0.25  & 0.02 \\ 
		& [0.41 ; 0.53]  (81) & [0.17 ; 0.32] (92) & [-0.06 ; 0.09]  (91)\\
 \rowcolor{Gray}  Pleyber-Christ	& 0.22 &  {0.10} 
 & 0.36 \\
\rowcolor{Gray}  & [0.16 ;  {0.27}
] (77) & [0.02 ; 0.17] (82) & [0.30 ; 0.43]  (80) \\
 \hline
\end{tabular}
\caption{Estimated auto-regressive coefficients $\beta_{ij}$ in (\ref{General_Model}) for the three weather stations plotted in Figure \ref{Stations}. 
The intervals represent the 95\% confidence intervals and the number between brackets corresponds to the 95\%  coverage probability.} 
\label{Tab MAR}
\end{table}

 Concerning the three atmospheric variables in $\Mb{F}_t$ (temperatures,   pressures at the sea level,  and  humidity) that drive our variability $\sigma_t$,  
 Table \ref{Parameter_Values} provides the respective estimated   regression coefficients from (\ref{eq: var esp}) with their associated $95\%$ confidence intervals and coverage probability. 
The values of $\hat{\theta}_j$ for $j=1,2,3$  are small but their $95\%$ confidence intervals do not contain zero. So, we keep these three explanatory variables in our rainfall generator.  
 
 \begin{table}[h]
\center
\begin{tabular}{|c|ccc|}
\hline
 Atmospheric variables         & Estimates & Confidence\  interval & Coverage\  probability \\
           \hline
           \hline
           
\rowcolor{Gray}   Regression intercept & $\hat{\theta}_0 = {30.626}$
 & [28.02 ; 32.32] & 80 \\
Temperature & $\hat{\theta}_1 =0.070$ &  [0.064; 0.076] &  {38}
 \\
\rowcolor{Gray}   Seal level pressure & $\hat{\theta}_2 =-0.034$ & [-0.037 ; -0.031] & 94 \\
Humidity & $\hat{\theta}_3 =0.028$  & [0.022; 0.036] & 94 \\
\hline
\end{tabular}
\caption{Estimated  regression coefficients in (\ref{eq: var esp})  and corresponding $95\%$ confidence intervals in function of our three explanatory 
atmospheric variables.}
\label{Parameter_Values}
\end{table}

\begin{figure}[H]
\center
\includegraphics[scale=0.445]{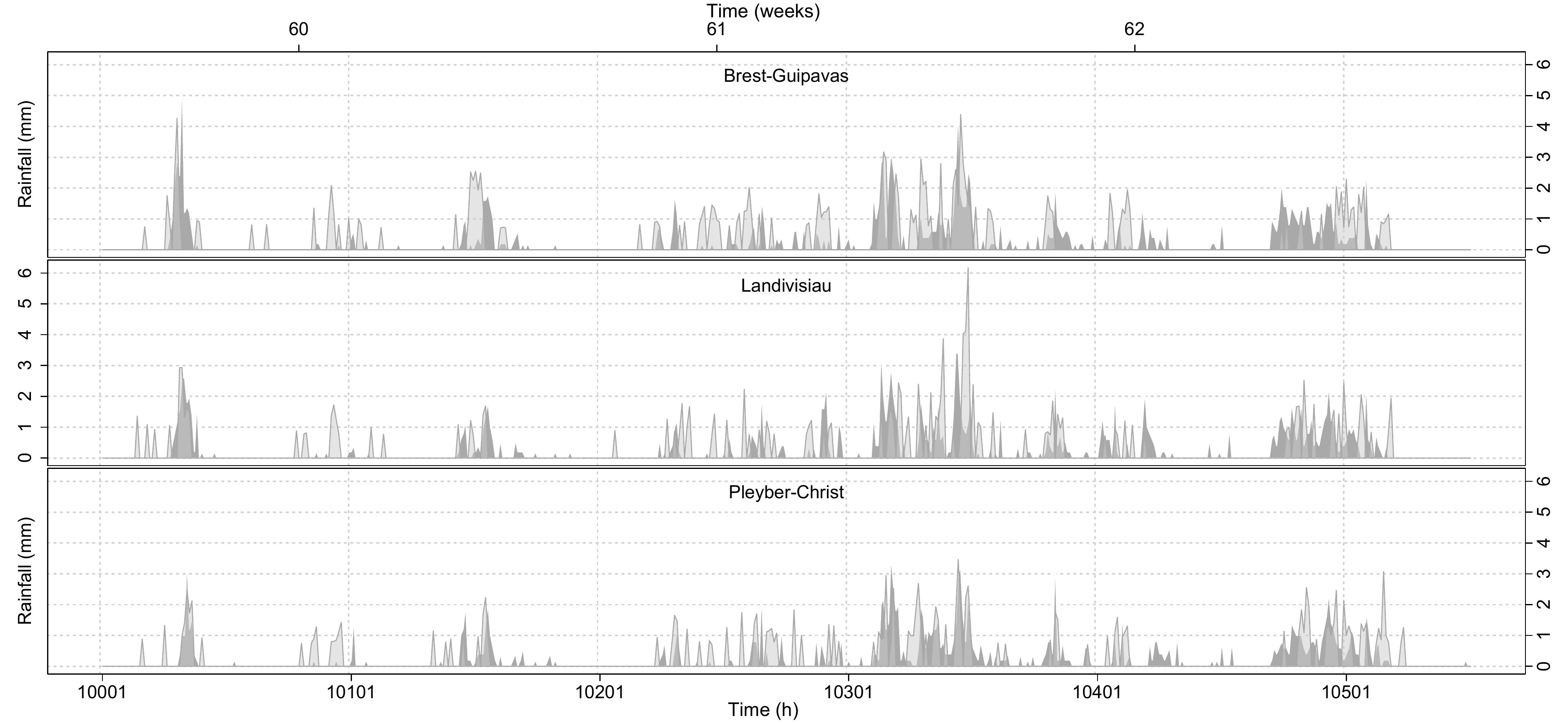}
\caption{
Each panel corresponds to one of our three weather stations shown in Figure \ref{Stations}. 
The dark  and light gray color represent  recorded precipitation and forecasted rainfall, respectively. 
Our predicted values are obtained on the validation period according to the scheme shown in Figure \ref{fig: design}. 
The regional factor $\sigma_t$ in (\ref{eq: var esp}) is the only quantity supposed to be known for the forecast. 
In particular, we don't use the precipitation recorded at time $t$ to forecast the following hour or the following week.
}
\label{Series_Outsample}
\end{figure}

To visualize the predictive capacity of our  model that has {been} 
fitted on the training set, we can generate synthetic hourly precipitation trajectories and compare them to   the one kept in our validation period. 
For each weather station shown in Figure \ref{Stations}, Figure \ref{Series_Outsample} compares    observed rainfall (dark gray) with the simulated one (light gray) during 500 hours of the  validation period, 
see Figure \ref{fig: design}. Note that we have chosen a relative short period for the sake of visibility. However, the conclusions that can be drawn here are true whatever the period considered in the validation set (figures available upon request).
 In this exercise, we  assume that the three atmospheric variables in $\Mb{F}_t$, temperature,   sea level  pressure  and  humidity,  are known at the regional scale.

\begin{figure}[h!]
\center
\includegraphics[scale=0.35]{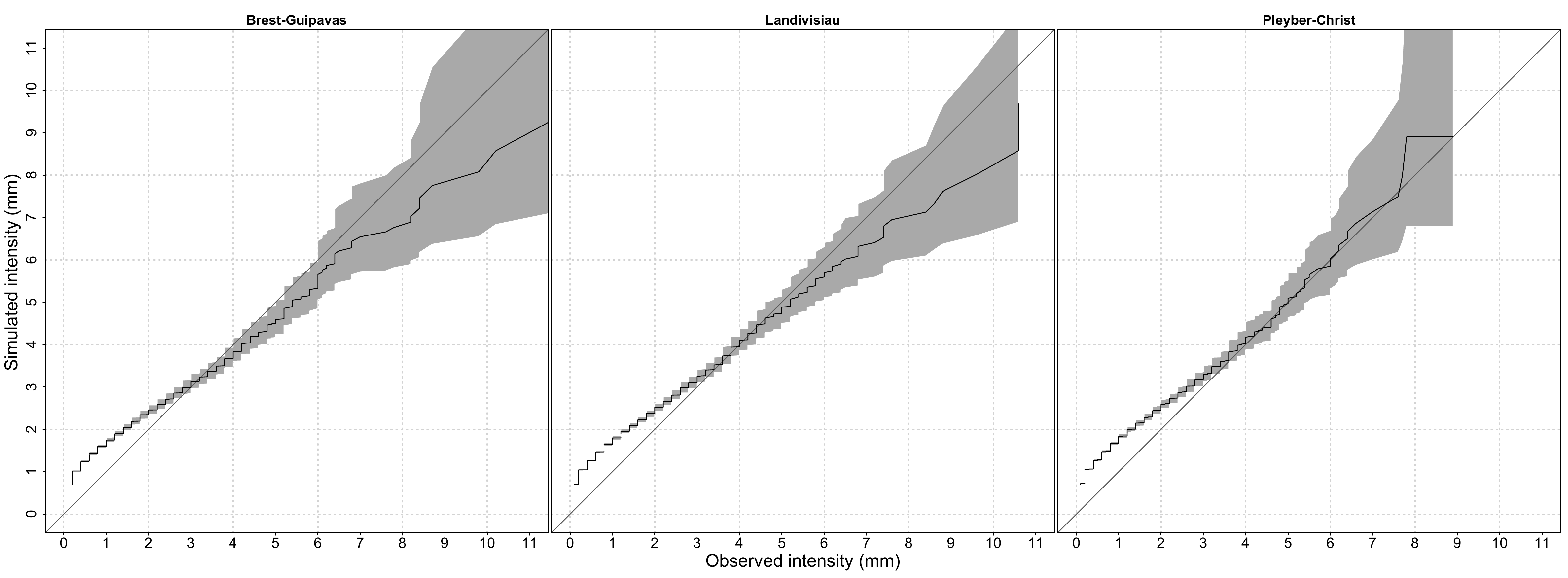}
\caption{Out-sample quantile-quantile plot between observed  rainfall amount (x-axis) and simulated one (y-axis) from model (\ref{General_Model}). 
Each  panel corresponds to one of our three weather stations shown in Figure \ref{Stations}. 
The gray color corresponds to the 98\% confidence interval and the solid line to the median.
}
\label{fig: qqplot1}
\end{figure}

\begin{figure}[h!]
\center
\includegraphics[scale=0.35]{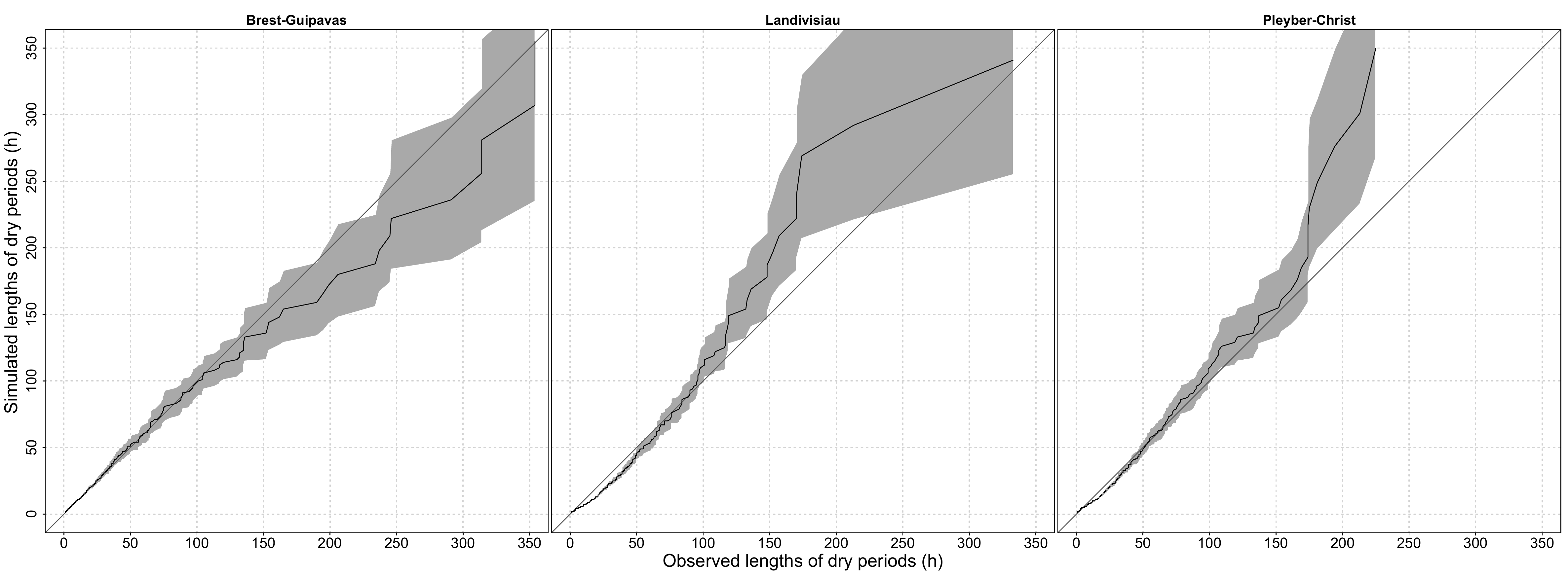}
\caption{Out-sample quantile-quantile plot between observed  {dry periods length} (x-axis) 
and simulated one (y-axis) from model (\ref{General_Model}). 
The gray color corresponds to the 98\% confidence interval and the solid line to the median.
}
\label{fig: qqplot2}
\end{figure}

The simulated rainfall   appear to    reproduce well the dynamical structure  of real precipitation in the sense that clusters of simulated rainfall seem to be temporally
 and spatially synchronized with the observations.
This indicates that combining  the  spatial factor  $\Mb{F}_t$ with a multivariate auto-regressive structure drives accurately the dynamics of the system. 
  Consequently,  our model,  conditionally on the common factor,  cannot only be used as a rainfall generator but it can also   carry out predictions. One can also  notice in Figure \ref{Series_Outsample} that real rainfall (left panel) have a jigsaw pattern. This is due to the recording device that is discrete in nature.  Recorded precipitation amounts  are added  by increment of $.2$ mm (the precision of the instrument). This mechanical feature is not built in our statistical model defined by (\ref{General_Model}) because we view this as  an undesirable characteristic of measured precipitation. 
This discrepancy explains the difference of ``granularity" between the two panels. 
This phenomenon is particularly important for small and medium precipitation amounts. 
To illustrate this, we show  the quantile-quantile plot between  recorded and simulated rainfall intensities in Figure \ref{fig: qqplot1}. 
The median and  the 98\% confidence interval in gray, obtained by parametric bootstrap (over 1000 simulated out-sample series), have been added.
Rainfall amounts above $3$ mm are well captured by our statistical model, this is particularly  true for extreme intensities. 
Precipitation under  $3$ mm  appear to be more difficult to reproduce and their intensities  are slightly overestimated. 
This may  {be}
 due to the nature of the recording process, the  jigsaw pattern due to the instrument precision,   and also to be the choice of our threshold $u$ in model (\ref{General_Model}). 
The latter was optimized ($u=0.7$ mm) to capture accurately  dry episodes. 
Figure \ref{fig: qqplot2} shows that this objective has been basically reached. 
Both long and short dry episodes appear to be simulated accurately by our statistical model, especially at the Brest-Guipavas station. 

\begin{figure}[h!]
\center
\includegraphics[scale=0.58]{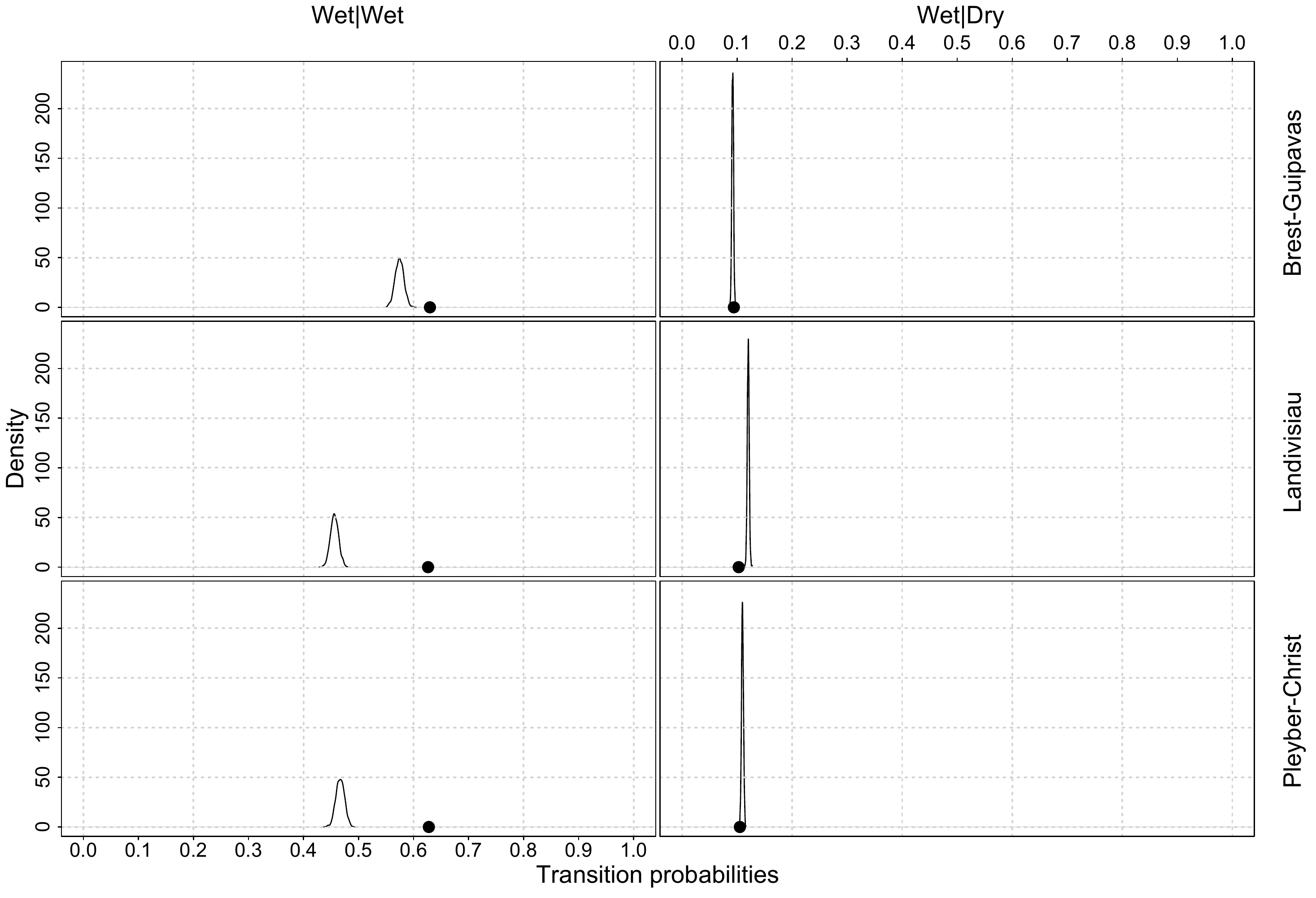}
\caption{ 
Each row represents a weather station. 
The left and right panels display the out-sample probability of having a wet hour given that the preceding hour was  wet   (left) and dry (right). The black points represent the observed estimates whereas the curves correspond to the density of estimates over 1000 out-sample simulated series.
}
\label{fig: trans}
\end{figure}

To go one step further in analyzing our predictive out-sample, 
the probability  of having a wet hour given that the preceding hour was also wet is displayed on the left panels of Figure \ref{fig: trans}, as well as the probability of moving from a dry hour to a wet one (right panels). 
Overall, these transition probabilities appear to give reasonable values, the black point corresponding to  the observed estimate and the density to the distribution from simulated out-samples. 
This is particularly true for the dry to wet transition. Let us note that the other transition probabilities stem directly from these due to the relations $\mathds{P}(\mbox{Wet}|\mbox{Wet})+\mathds{P}(\mbox{Dry}|\mbox{Wet})=1$ and $\mathds{P}(\mbox{Wet}|\mbox{Dry})+\mathds{P}(\mbox{Dry}|\mbox{Dry})=1$.

\begin{figure}[h!]
\center
\includegraphics[scale=0.65]{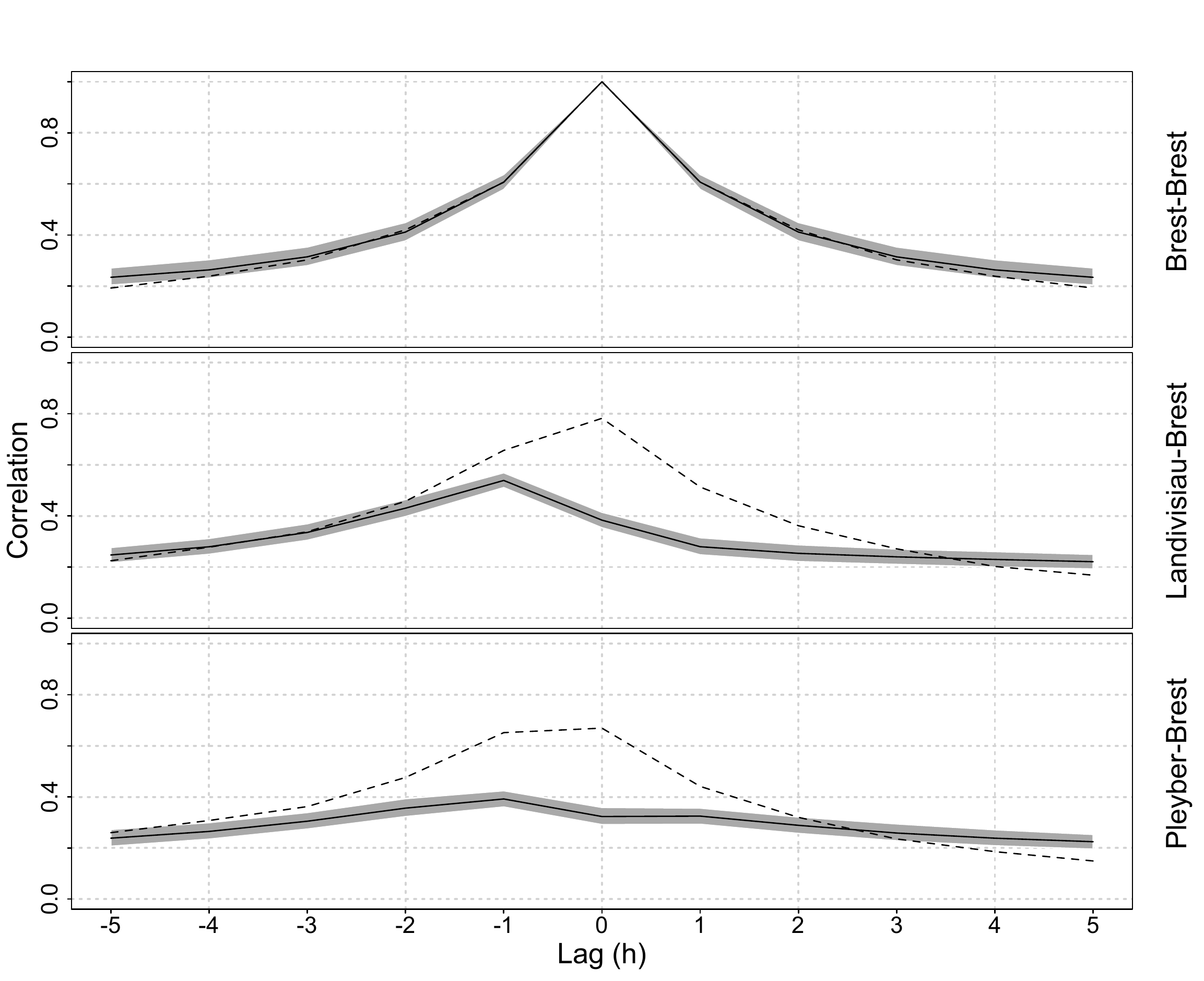}
\caption{These panels correspond to out-sample correlations at different lags. On top, the auto-correlation at Brest-Guipavas. On the middle, the cross-correlation between Landivisiau and Brest-Guipavas and on bottom the cross-correlation between Pleyber-Christ and Brest-Guipavas. The gray color corresponds to the 98\% confidence interval, the solid line to the median and the dashed one to the observed correlations.
}
\label{fig: correl}
\end{figure}

Concerning the temporal memory, the top panel of Figure \ref{fig: correl} shows the out-sample auto-correlations at different time lags of one hour for the Brest-Guipavas station. 
This graph indicates that this local short term persistence (from one to five hours) is very well reproduced.  
The lower two panels focus on the out-sample cross-correlations between two pairs of stations, Landivisiau-Brest and Pleyber-Brest, respectively. 
These correlations are more difficult to reproduce: the observed ones (dotted lines) are above the ones computed on our out-sample. However, we capture the temporal asymmetry. The understimation of correlations at lag $-1$, $0$ and $1$ stems partly from the fact that the $\{ \epsilon_{m,t} \}_{m=1, \dots, M}$ are i.i.d. This point will be further discussed in conclusion.

\begin{figure}[h!]
\center
\includegraphics[scale=0.455]{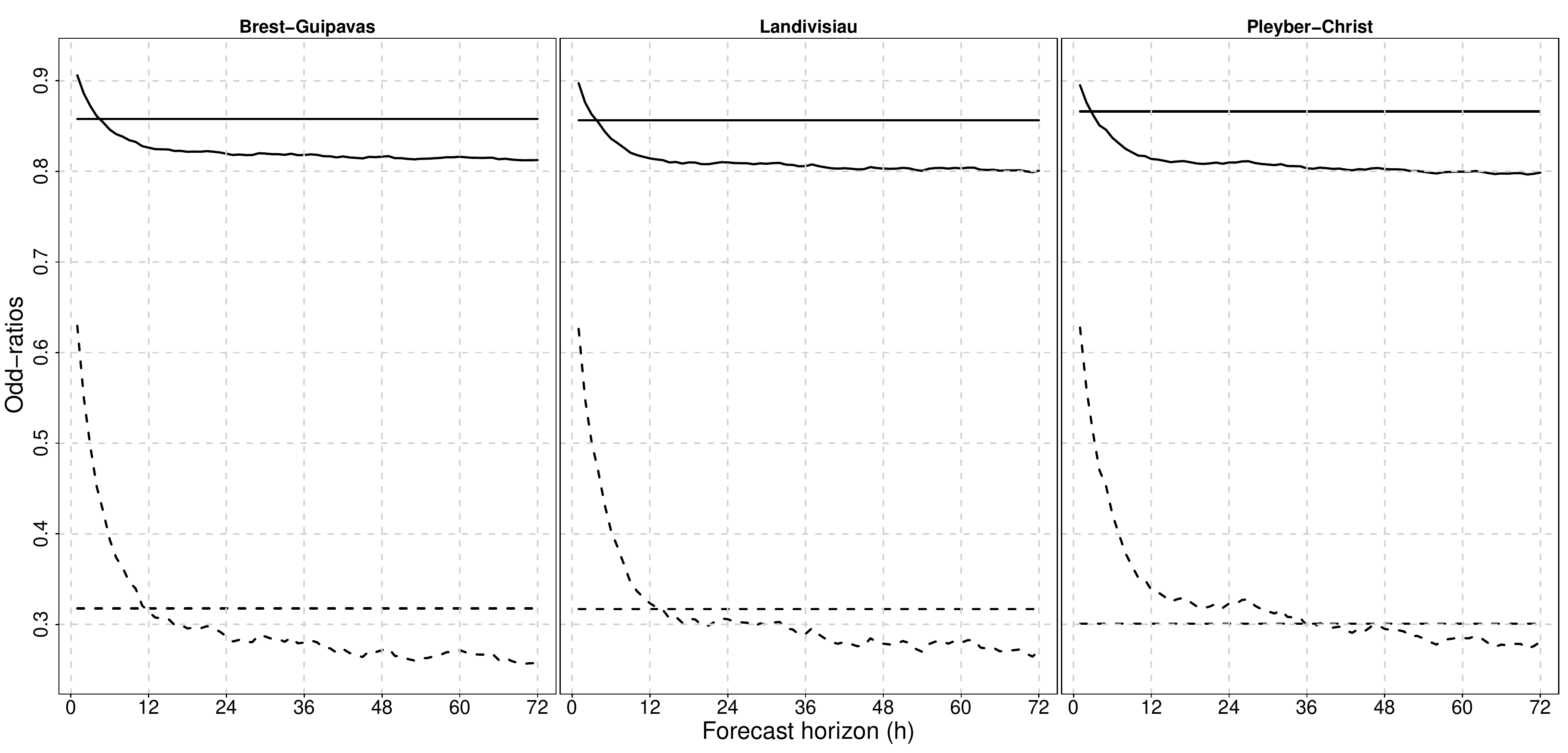}
\caption{Each  panel corresponds to one of our three weather stations shown in Figure \ref{Stations}. In solid are displayed the estimated probabilities to predict a dry hour while this hour is effectively dry. The constant line corresponds to the estimate obtained with our model. The curve corresponds to the model predicting for a given horizon the same value as now. In dashed, the same concerns the probabilities to predict a wet hour while this hour is effectively wet.}
\label{fig: Odd_Ratios}
\end{figure}

In order to analyze the prediction ability of our model, we have computed the following odd-ratios: the out-sample probability of predicting a dry hour while this hour is indeed dry and the out-sample probability of predicting a wet hour while this hour is indeed wet. In order to carry out a comparison, we have considered as a reference the simple model that gives as a prediction for a future horizon the value observed now. In Figure \ref{fig: Odd_Ratios}, we observe that from a certain value of the horizon, our model outperforms the naive prediction scheme. In some sense, this test allows to see approximately at which moment the frailty factor takes over the contagion term.

In our example, the frailty factor plays an important role in reproducing accurately  temporal dynamics,  rainfall intensity and dry period persistence.  
To test this (figures available upon request), our model was fitted without the frailty component and the aforementioned features were not adequately  reproduced in such an instance.
We have also fitted our model without the contagion term (figures available upon request). The performance is quite good, although the persistence of wet periods seems to be underestimated.

We would like to emphasize that figures \ref{Series_Outsample}-\ref{fig: Odd_Ratios} have been obtained on a validation set, totally different from  the training one used to fit the model. This is reassuring and, albeit a few adjustments or extensions,  we can expect that the basic ideas developed here could  be relevant  for other regions.   These aspects are discussed in our conclusion.
%

Finally, we have also applied our model on daily precipitations measured at the same stations (see Figure \ref{Stations}). Our model's performance is at least as good as in the hourly case. Corresponding figures are available upon request. To give perspective, we compared our model with the standard model by \cite{Wilks98}. Our model gives similar results concerning the general statistical properties (distributions of intensity and dry periods lengths, transition probabilities, cross-correlations, \dots). Due to the absence of covariates, Wilks' model does not capture the dynamic structure, and especially the synchronicity between observed and simulated rainfalls. 
As already mentioned, these features are well reproduced by our model.

\newpage

\section{Concluding remarks}
\label{Chaprainfall_Sec_Conclusion}
Our main  objective was to show  that a "simple" multivariate auto-regressive model with a heteroscedastic variance driven by a few well-chosen atmospheric covariates can provide an interesting blueprint to generate dry episodes, medium and heavy rainfall.
From there, it is easy to extend this work by adding or modifying a few elements of our model. 
For example, we could imagine to replace the spatially independent  random Gaussian noise $\epsilon_{m,t}$ by a multivariate Gaussian vector with a covariance matrix that could represent some  spatial dependence among the different weather stations. One can even think about a multivariate student or elliptical distribution in order to provide heavier tails. 
This may be needed in regions with very heavy rainfall. Our northern Brittany example is known for frequent rainfall episodes but with rather low intensities, say compared to the South of France. 

Another possible road for improvements would be to replace the linearity assumption in our   regression model (\ref{eq: var esp}) by a non-parametric relationship (like a  GAM,  \cite[e.g.,][]{Serinaldi14}) that will allow to capture some non-linear behavior. 
Although it was not  {really}
needed for our Brittany example, it is  likely that the link between rainfall variability and temperatures could {be}
 more complex than our  linear model for other regions. 

It is also true that our multi-site statistical model is not a pure  precipitation weather generator 
in the sense that we need more than  precipitation data to fit and run our model. A few atmospheric covariates are necessary  to drive  our rainfall variability (for simulation of temperature time series, see e.g. \cite{huong2009multidimensional} and \cite{hoang2011estimation}).  
This limitation could be viewed as an advantage in the context of climate change studies. 
By letting our spatially averaged atmospheric covariates being driven by a numerical model, we could explore how precipitation change under different forcings.
This is obviously related to downscaling themes and it would be interesting to pursue this path in future work. 
In this context, selecting appropriate  covariates could be even more relevant, especially for large regions where scale indices like the North Atlantic Oscillation (NAO) Index or the El Nino Southern Oscillation (ENSO) Index could be useful.

One possible drawback of our approach may reside in our small number of sites. 
Modeling  tens or even hundreds of stations, instead of three, will lead to computationally difficulties because of the $M \times M$ size of the matrix $\bf B$. 
This could be solved by imposing a parametric structure on the $\beta_{ij}$, for example they could 
decrease
with the distance with respect to other stations or even set to zero for far apart stations. 
Finding a suitable distance is not trivial and should depend on orographic and other physical features. 

To conclude, there are many different ways to fine tune and extend  our approach. 
Basically, this will strongly depend on the application at hand and more research is needed to clearly see if this framework can be generalized.

\section*{Acknowledgements}
We thank Pierre Ribereau, Christian Y. Robert
and the participants at the 12th International Meeting on Statistical Climatology and the ISI World Statistics Congress 2013 for their useful comments and suggestions.
The authors have been financially supported by the  different projects:    ANR  McSim, MIRACCLE-GICC, LEFE-Multirisk  and Extremoscope. Erwan Koch would like to thank RiskLab at ETH Zurich and the Swiss Finance Institute for financial support.

\newpage
\appendix
\section{Likelihood computation}

\begin{proof}
We consider that the common factor's path $\Mb{F}_t \ (t=1, \dots, T)$ is given. Therefore we omit $\Mb{F}_t$ in the following. Let us denote by $I_{t-1}$ the information available at time $t-1$, \\
$ I_{t-1} = (P_{1,t-1}, \dots, P_{M,t-1}, P_{1,t-2}, \dots, P_{M,t-2}, \dots)$ and by $h$ the density function.
We can write by using a sequential argument that 
\Beq
\label{Eq_Recurrence_Conditional}
h(P_{1,t}, \dots, P_{M,t}\ ; t \in 1, \dots, T) = \prod_{t=2}^T  h(P_{1,t}, \dots, P_{M,t}  |  P_{1,t-1}, \dots, P_{M,t-1}).
\Eeq

Since the $\epsilon_{m,t}, m=1, \dots, M$ are i.i.d. (conditionally on $\Mb{F}_t$),
the variables
$${\bf B}_m^{'} {\bf P}_{t-1} + \epsilon_{m,t}, m=1, \dots, M$$ are independent. 
Thus it is the same for the variables $${\bf I}_{{\bf B}_m^{'} {\bf P}_{t-1} + \epsilon_{m,t} >u}, m=1, \dots, M $$
and therefore for the product
$$ P_{m,t}= \left( {\bf B}_m^{'} {\bf P}_{t-1} + \epsilon_{m,t} \right) {\bf I}_{{\bf B}_m^{'} {\bf P}_{t-1} + \epsilon_{m,t} >u}, m=1, \dots, M.$$
Therefore the measurement equation given by model (\ref{General_Model}) gives that conditionally on $(I_{t-1})$, the variables $P_{1,t}, \dots, P_{M,t}$ are independent, yielding
\Beq
\label{Eq_Conditional_Independence}
h(P_{1,t}, \dots, P_{M,t}  |  P_{1,t-1}, \dots, P_{M,t-1}) = \prod_{m=1}^M h(P_{m,t}  |  P_{1,t-1}, \dots, P_{M,t-1}).
\Eeq
Combining Equations \eqref{Eq_Recurrence_Conditional} and \eqref{Eq_Conditional_Independence}, we obtain
\Beq
\label{Paper_WeatherGenerator_Eq_Density}
h(P_{1,t}, \dots, P_{M,t}; t \in 1, \dots, T) = \prod_{t=2}^T  \prod_{m=1}^M h(P_{m,t}  |  P_{1,t-1}, \dots, P_{M,t-1}).
\Eeq 

Let us recall that
$$ P_{m,t}=
\left \{
\begin{array}{cc}
{\bf B}_m^{'} {\bf P}_{t-1} + \epsilon_{m,t} & \mbox{ if } {\bf B}_m^{'} {\bf P}_{t-1} + \epsilon_{m,t} \geq u, \\
0 & \mbox{ if } {\bf B}_m^{'} {\bf P}_{t-1} + \epsilon_{m,t} < u.
\end{array}
\right.
$$
Due to this threshold mechanism, $P_{m,t}$ is a mixture of a discrete random variable and a continuous one. Let us denote by $h_d$ the density of the discrete part and by $h_c$ the density of the continuous one. We have
$$ hc(P_{m,t}  |  P_{1,t-1}, \dots, P_{M,t-1}) =  \frac{1}{\sigma_t} \phi \left( \frac{P_{m,t} -{\bf B}_m^{'} {\bf P}_{t-1}}{\sigma_t} \right)
$$
and
\begin{align*}
hd(P_{m,t}  |  P_{1,t-1}, \dots, P_{M,t-1}) &= \mathds{P} \left[ P_{m,t}=0 \ | P_{1,t-1}, \dots, P_{M,t-1} \right]
\\& = \mathds{P} \left[ {\bf B}_m^{'} {\bf P}_{t-1} + \epsilon_{m,t} < u \right]
\\& = \mathds{P} \left[ \epsilon_{m,t} < u -{\bf B}_m^{'} {\bf P}_{t-1} \right]
\\& = \Phi \left( \frac{u - {\bf B}_m^{'} {\bf P}_{t-1} }{\sigma_t} \right).
\end{align*}

Using Equation \eqref{Paper_WeatherGenerator_Eq_Density}, the log-density is written
\begin{align*}
& \ln \left[ h(P_{1,t}, \dots, P_{M,t}, t=1, \dots, T) \right]
\\& = \sum_{t=2}^T \sum_{m=1}^M 
\ln \left[ hc(P_{m,t}  |  P_{1,t-1}, \dots, P_{M,t-1}) \ {\bf I}_{ \{ P_{m,t} \geq u  \} } + hd(P_{m,t}  |  P_{1,t-1}, \dots, P_{M,t-1}) \ {\bf I}_{ \{ P_{m,t} = 0 \} } \right]
\\& = \sum_{t=2}^T \sum_{m=1}^M {\bf I}_{ \{ P_{m,t} \geq u  \} } \ln \left[ hc(P_{m,t}  |  P_{1,t-1}, \dots, P_{M,t-1}) \right] + {\bf I}_{ \{ P_{m,t} = 0\} }\  \ln \left[ hd(P_{m,t}  |  P_{1,t-1}, \dots, P_{M,t-1}) \right].
\end{align*}

That finally yields the log-likelihood function
\begin{align*}
& L_u({\bf B}, \theta_0, \dots, \theta_d) 
\\& = \sum_{t=2}^T \sum_{m=1}^M \left\{ 
{\bf I}_{ \{ P_{m,t} \geq u  \} } \ln \Bigg[ \frac{1}{\sigma_t} \phi \left( \frac{P_{m,t} - {\bf B}_m^{'} {\bf P}_{t-1}}{\sigma_t} \right) \Bigg] + {\bf I}_{ \{P_{m,t} = 0 \} } \ln \Bigg[ \Phi \left( \frac{u - {\bf B}_m^{'} {\bf P}_{t-1}}{\sigma_t} \right)  \Bigg]
\right\},
\end{align*}
completing the proof.
\end{proof}

\newpage
\bibliographystyle{apalike}
\bibliography{Biblio_Papier_Naveau-1}

\begin{thebibliography}{}

\bibitem[Ailliot et~al., 2014]{Ailliot14}
Ailliot, P., Allard, D., Monbet, V., and Naveau, P. (2014).
\newblock Stochastic weather generators: an overview of weather type models.
\newblock {\em Submitted}.

\bibitem[Ailliot et~al., 2009]{Ailliot09}
Ailliot, P., Thompson, C., and Thomson, P. (2009).
\newblock Space time modeling of precipitation using a hidden {M}arkov model
  and censored {G}aussian distributions.
\newblock {\em Journal of the Royal Statistical Society, Series C (Applied
  Statistics)}, 58(3):405--426.

\bibitem[Allard, 2012]{Allard12}
Allard, D. (2012).
\newblock Modeling spatial and spatio-temporal non {G}aussian processes.
\newblock In Porcu, E., Montero, J.-M., and Schlather, M., editors, {\em
  Advances and Challenges in Space-time Modelling of Natural Events}, Lecture
  Notes in Statistics, pages 141--164. Springer Berlin Heidelberg.

\bibitem[Allard and Bourotte, 2013]{Allard13}
Allard, D. and Bourotte, M. (2013).
\newblock Disaggregating daily precipitations into hourly values with a
  transformed censored latent {G}aussian process.
\newblock {\em Stochastic Environmental Research and Risk Assessment},
  Submitted.

\bibitem[Azizpour et~al., 2008]{azizpour2008self}
Azizpour, S., Giesecke, K., et~al. (2008).
\newblock Self-exciting corporate defaults: contagion vs. frailty.
\newblock {\em Manuscript, Stanford University}.

\bibitem[Bailey, 1953]{bailey1953total}
Bailey, N.~T. (1953).
\newblock The total size of a general stochastic epidemic.
\newblock {\em Biometrika}, 40(2):177--185.

\bibitem[Bailey, 1957]{bailey1957mathematical}
Bailey, N.~T. (1957).
\newblock {\em The mathematical theory of epidemics}.
\newblock Griffin London.

\bibitem[Davis and Lo, 2001]{davis2001modelling}
Davis, M. and Lo, V. (2001).
\newblock Modelling default correlation in bond portfolios.
\newblock {\em Mastering risk}, 2:141--151.

\bibitem[Davis et~al., 2014]{Davis14}
Davis, R.~A., Zang, P., and Zheng, T. (2014).
\newblock Sparse vector autoregressive modeling.
\newblock {\em Submitted}.

\bibitem[Flecher et~al., 2010]{Flecher10}
Flecher, C., Naveau, P., Allard, D., and Brisson, N. (2010).
\newblock A stochastic daily weather generator for skewed data.
\newblock {\em Water Resources Research}, 46(7).

\bibitem[Frey and Backhaus, 2003]{frey2003interacting}
Frey, R. and Backhaus, J. (2003).
\newblock Interacting defaults and counterparty risk: a {M}arkovian approach.
\newblock {\em University of Leipzig, Leipzig, Germany}.

\bibitem[Furrer and Katz, 2008]{Furrer08}
Furrer, E. and Katz, R. (2008).
\newblock Improving the simulation of extreme precipitation events by
  stochastic weather generators.
\newblock {\em Water Resources Research}, 44(12).

\bibitem[Gabriel and Neumann, 1962]{Gabriel62}
Gabriel, K. and Neumann, J. (1962).
\newblock A {M}arkov chain model for rainfall occurrence at {T}el {A}viv.
\newblock {\em Quarterly Journal of the Royal Meteorological Society},
  88:90--95.

\bibitem[Gagliardini and Gouri{\'e}roux, 2013]{gagliardini2013correlated}
Gagliardini, P. and Gouri{\'e}roux, C. (2013).
\newblock Correlated risks vs contagion in stochastic transition models.
\newblock {\em Journal of Economic Dynamics and Control}, 37(11):2241--2269.

\bibitem[Giesecke and Weber, 2004]{giesecke2004cyclical}
Giesecke, K. and Weber, S. (2004).
\newblock Cyclical correlations, credit contagion, and portfolio losses.
\newblock {\em Journal of Banking \& Finance}, 28(12):3009--3036.

\bibitem[Giesecke and Weber, 2006]{giesecke2006credit}
Giesecke, K. and Weber, S. (2006).
\newblock Credit contagion and aggregate losses.
\newblock {\em Journal of Economic Dynamics and Control}, 30(5):741--767.

\bibitem[Grimaldi et~al., 2005]{Grimaldi05}
Grimaldi, S., Serinaldi, F., and Tallerini, C. (2005).
\newblock Multivariate linear parametric models applied to daily rainfall time
  series.
\newblock {\em Advances in Geosciences}, 2(2):87--92.

\bibitem[Hoang et~al., 2011]{hoang2011estimation}
Hoang, T., Dacunha-Castelle, D., and Benmenzer, G. (2011).
\newblock Estimation of a diffusion model with trends taking into account the
  extremes. {A}pplication to temperature in {F}rance.
\newblock {\em Environmetrics}, 22(3):464--479.

\bibitem[Huong~Hoang et~al., 2009]{huong2009multidimensional}
Huong~Hoang, T., Parey, S., and Dacunha-Castelle, D. (2009).
\newblock Multidimensional trends: The example of temperature.
\newblock {\em The European Physical Journal-Special Topics}, 174(1):113--124.

\bibitem[Katz, 1977]{Katz77}
Katz, R. (1977).
\newblock Precipitation as a chain-dependant process.
\newblock {\em Journal of Applied Meteorology}, 16:671--676.

\bibitem[Kendall, 1956]{kendall1956deterministic}
Kendall, D.~G. (1956).
\newblock Deterministic and stochastic epidemics in closed populations.
\newblock In {\em Proceedings of the third {B}erkeley Symposium on Mathematical
  Statistics and Probability}, volume~4, pages 149--165.

\bibitem[Kleiber et~al., 2012]{Kleiber12}
Kleiber, W., Katz, R., and Rajagopalan, B. (2012).
\newblock Daily spatiotemporal precipitation simulation using latent and
  transformed {G}aussian processes.
\newblock {\em Water Ressources Research}, 48(1).

\bibitem[Lennartsson et~al., 2008]{lennartsson08}
Lennartsson, J., Baxevani, A., and Chen, D. (2008).
\newblock Modelling precipitation in {S}weden using multiple step {M}arkov
  chains and a composite model.
\newblock {\em Journal of Hydrology}, 363(1):42--59.

\bibitem[Maraun et~al., 2010]{maraun2010}
Maraun, D., Wetterhall, F., Ireson, A., Chandler, R., Kendon, E., Widmann, M.,
  Brienen, S., Rust, H., Sauter, T., Theme{\ss}l, M., et~al. (2010).
\newblock Precipitation downscaling under climate change: Recent developments
  to bridge the gap between dynamical models and the end user.
\newblock {\em Reviews of Geophysics}, 48(3).

\bibitem[Pawitan, 2001]{Yudi01}
Pawitan, Y. (2001).
\newblock {\em All Likelihood: Statistical Modelling and Inference Using
  Likelihood}.
\newblock Oxford University Press.

\bibitem[Richardson, 1981]{Richardson81}
Richardson, C. (1981).
\newblock Stochastic simulation of daily precipitation, temperature, and solar
  radiation.
\newblock {\em Water Resources Research}, 17(1):182--190.

\bibitem[Serinaldi and Kilsby, 2014]{Serinaldi14}
Serinaldi, F. and Kilsby, C.~G. (2014).
\newblock Simulating daily rainfall fields over large areas for collective risk
  estimation.
\newblock {\em Journal of Hydrology}, 512:285--302.

\bibitem[Srikanthan et~al., 2001]{srikanthan01}
Srikanthan, R., McMahon, T., et~al. (2001).
\newblock Stochastic generation of annual, monthly and daily climate data: A
  review.
\newblock {\em Hydrology and Earth System Sciences Discussions}, 5(4):653--670.

\bibitem[Vrac and Naveau, 2007]{Vrac07a}
Vrac, M. and Naveau, P. (2007).
\newblock Stochastic downscaling of precipitation: from dry events to heavy
  rainfalls.
\newblock {\em Water Resources Research}, 43(7).

\bibitem[Vrac et~al., 2007]{Vrac07}
Vrac, M., Stein, M., and Hayhoe, K. (2007).
\newblock Statistical downscaling of precipitation through non homogeneous
  stochastic weather typing.
\newblock {\em Climate Research}, 34:169--184.

\bibitem[Wilks, 1998]{Wilks98}
Wilks, D. (1998).
\newblock Multisite generalization of a daily stochastic precipitation
  generation model.
\newblock {\em Journal of Hydrology}, 210(1):178--191.

\bibitem[Wilks, 2010]{wilks2010use}
Wilks, D. (2010).
\newblock Use of stochastic weather generators for precipitation downscaling.
\newblock {\em Wiley Interdisciplinary Reviews: Climate Change}, 1(6):898--907.

\bibitem[Wilks, 2012]{wilks2012stochastic}
Wilks, D. (2012).
\newblock Stochastic weather generators for climate-change downscaling, part
  ii: multivariable and spatially coherent multisite downscaling.
\newblock {\em Wiley Interdisciplinary Reviews: Climate Change}, 3(3):267--278.

\bibitem[Wilks and Wilby, 1999]{WilksWilby99}
Wilks, D. and Wilby, R. (1999).
\newblock The weather generation game : a review of stochastic weather models.
\newblock {\em Progress in Physical Geography}, 23(3):329--357.

\end{thebibliography}
\end{document}